\documentclass[preprint,nofootinbib,superscriptaddress,amsmath,amssymb,UTF8]{revtex4}
\usepackage{amsmath,mathrsfs,amsbsy,color,graphicx,bm,amsthm,amsfonts}
\usepackage{bbm}
\usepackage{times}
\usepackage{units}
\usepackage{dcolumn}
\usepackage{graphicx}
\usepackage{epsfig}
\usepackage{epstopdf}
\usepackage[colorlinks,linkcolor=red,anchorcolor=green,citecolor=blue,CJKbookmarks=True]{hyperref}
\DeclareMathSymbol{\shortminus}{\mathbin}{AMSa}{"39}
\usepackage{mathrsfs}
\usepackage{braket}
\usepackage{amssymb}
\usepackage{txfonts}
\usepackage{float}
\usepackage{enumitem}
\usepackage{multirow}
\usepackage{orcidlink}
\usepackage{upgreek}

\begin{document}
\title{Bayesian Inference of Primary Hair from Black Hole Shadows in Generalized Proca Theory}
\author{Jiao-Zhen She}
\affiliation{School of Electrical and Information Engineering, Tongling University, Tongling 244061, China}

\author{Shu-Min Wu\orcidlink{0000-0001-7392-1327}}
\email[]{smwu@lnnu.edu.cn (corresponding author)} 
\affiliation{Department of Physics, Liaoning Normal University, Dalian 116029, China}

\begin{abstract}
We investigate black hole shadows and observational constraints on primary vector hair in an exact rotating solution of generalized Proca theory.
Selecting a coupling slice yields a simple, one-parameter Kerr deformation that is noncircular and depends on spin and latitude.
The deformation vanishes in the nonrotating limit, while the vector field remains nontrivial.
We determine the domain admitting a regular outer horizon and construct shadow and lensing images by numerical ray tracing.
Ray-traced images show a modest enlargement of the shadow and changes in the surrounding lensing pattern.
Bayesian inference using the shadow area-equivalent radius and axis ratio yields joint constraints on hair, spin, and viewing inclination from Event Horizon Telescope observations of M87* and Sgr A*.
The M87* posterior favors relatively low spin and exhibits a pronounced spin--hair anticorrelation, while Sgr A* yields upper limits on spin and hair with its inclination remaining essentially unconstrained.
These results connect the optical signatures of primary vector hair to quantitative horizon-scale constraints.
\end{abstract}

\maketitle

\section{Introduction}
The Event Horizon Telescope (EHT) observations of M87* and Sgr A* have opened a new window for testing gravity in the strong-field regime \cite{EventHorizonTelescope:2019ggy,EventHorizonTelescope:2022wkp,EHT:2025persistent,EventHorizonTelescope:2025vum,Lu:2023ring,Saurabh:2026jetbase}.
Black hole shadows and related imaging observables provide valuable information about the underlying spacetime geometry and its possible departures from the Kerr geometry \cite{Wei:2013emdashadow,Chen:2022scf,Wang:2021ara,Hou:2022kerrmelvin,Zhang:2024thickimages,Zhang:2024jrw,Wang:2025kbrgeodesics,Zeng:2025kbroptics,Wang:2025btn,Qin:2025braneworld,Qin:2025kzpolarized,Chatterjee:2025extraction,Cunha:2015yba,He:2025kzimages,Huang:2025parityoddshadows,Meng:2025hairimages}.
In modified gravity theories, additional degrees of freedom can modify black hole shadows and gravitational lensing, offering potential observational signatures of black hole hair \cite{Wan:2025horndeskiimages}.
However, extracting information about additional gravitational degrees of freedom from shadow observations is challenging, as the observed image also depends on black hole spin and viewing inclination \cite{Wang:2022mjo,Faggert:2025asymmetry,Bernshteyn:2026ringspin}.
This motivates studying a concrete black hole solution with primary hair to determine how the hair modifies observable shadow features and what constraints can be obtained from current observations.

Generalized Proca theory provides a vector--tensor counterpart of Horndeski scalar--tensor theory \cite{Horndeski:1974wa,Deffayet:2011gz,Kobayashi:2011nu}, with derivative interactions that retain second-order field equations and three propagating vector polarizations \cite{Heisenberg:2014rta,Tasinato:2014eka,Allys:2015sht,BeltranJimenez:2016rff,Allys:2016jaq,Dong:2024procapolarization}.
Its nonminimal vector couplings admit black holes with primary hair, introducing integration data independent of mass and angular momentum \cite{Heisenberg:2017xda}.
Recent developments span models derived from regularized Gauss--Bonnet and Lovelock invariants and exact rotating solutions with noncircular geometries \cite{Charmousis:2025proca,Fernandes:2025lovelockproca,Fernandes:2026primary}.
Within one branch of this rotating solution family, a constrained subfamily admits an equilibrium thermodynamic description through second order in rotation \cite{Liu:2026constrainedhair}; complementary studies of black hole thermodynamics classify equilibrium branches through their thermodynamic topology \cite{Wei:2022topology,Wei:2022dzw,Wei:2024universal,Wu:2024rmv,Liu:2025rotatingtopology,Liu:2025noveltopology,Liu:2025acceleratingtopology,Jeon:2025topological,Wu:2025extendedtopology}.
Here we investigate the optical properties of another branch, whose hair-induced deformation depends on spin and latitude and vanishes in the nonrotating limit \cite{Fernandes:2026primary}.
This rotation-activated behavior intertwines the optical effects of hair, spin, and viewing inclination, while the existence of a regular outer horizon restricts the allowed spin--hair combinations.
How well shadow observations can distinguish these effects and constrain primary hair is the physical question motivating this study.

Investigating the observational signatures of such rotating black holes requires a quantitative connection between their spacetime geometry and observable shadow features.
For the noncircular geometry considered here, numerical ray tracing allows us to determine the shadow boundary and gravitational lensing structure by integrating the photon equations.
Extensive studies have demonstrated the effectiveness of numerical approaches in exploring black hole shadows and optical phenomena across a wide range of spacetime geometries \cite{Cunha:2019dwb,Wang:2017qhh,Hu:2020usx,Zhong:2021mty,Garnier:2023lph,Huang:2024wpj,Zhang:2025xnl,Konoplya:2025bte,Gao:2026xdt,Liu:2024lbi,Liu:2024lve,Liu:2024soc,Liu:2024iec,Liu:2025lwj,Sharma:2025mahakala,Guo:2022multiplephoton,Hu:2025tiltedinner}.
To constrain additional gravitational parameters, the predicted shadow observables must be compared quantitatively with observational measurements.
Liu et al. connected numerical shadow modeling with Bayesian parameter estimation \cite{Liu:2025wwq}, combining adaptive backward ray tracing with Schwarzschild calibration to infer horizon-scale magnetic fields in the Kerr--Bertotti--Robinson spacetime \cite{Podolsky:2025KBR}.
This observable-based inference strategy provides a suitable methodological foundation for investigating primary vector hair in the exact noncircular rotating geometry considered here.

In this work, we select a coupling slice of the exact rotating solution family to obtain a one-parameter primary-hair deformation of Kerr and numerically determine its horizon-admitting domain.
We then use ray-traced shadow and lensing images to examine how this deformation affects photon capture and the surrounding optical structure.
The shadow area-equivalent radius and axis ratio provide a quantitative description of these effects and define a numerical map from spin, hair, and inclination to observables.
Building on the inference strategy of Liu et al. \cite{Liu:2025wwq}, we use this map to obtain joint posterior distributions from the M87* and Sgr A* measurements.
The analysis identifies the parameter combinations favored by each source, resolves their spin--hair correlations, and assesses the information available about inclination.
Comparison with the adopted priors further establishes the role of the horizon-admitting domain and the sampled hair range in the inferred constraints.
The remainder of this paper is organized as follows.
Section~\ref{sec:theory} introduces the theory, specifies the coupling slice of the exact rotating solution family used to obtain our one-parameter primary-hair geometry, and determines its horizon-admitting parameter domain.
Section~\ref{sec:forward} uses numerical ray tracing to investigate the effects of primary hair on black hole shadows, gravitational lensing, and shadow observables.
Section~\ref{sec3} gives the Bayesian constraints from M87* and Sgr A*, and Sec.~\ref{sec:conclusion} concludes with a discussion of future directions.

\section{Theoretical background}\label{sec:theory}

\subsection{Generalized Proca theory and exact rotating solutions}

Exact rotating black holes are rare outside vacuum general relativity \cite{Filippini:2017kov}.
Even after stationarity and axial symmetry are imposed, the field equations remain a coupled system of nonlinear partial differential equations, and most studies therefore assume circularity or rely on perturbative and numerical constructions \cite{Babichev:2025szb,Candan:2025fbl,Cheng:2025hdw,Kleihaus:2026rev,Liu:2022dcn,Deng:2025uvp,Liu:2026modmaxmodes}.
The solution studied here avoids both restrictions: it is exact, asymptotically flat, and noncircular, and it contains an integration constant associated with the vector field that is independent of the ADM mass and angular momentum \cite{Fernandes:2026primary}.
Its theoretical origin can be traced from Generalized Proca theory to a class of four-dimensional vector-tensor models inspired by regularized Lovelock gravity.

The derivative interactions in Generalized Proca theory preserve the three propagating degrees of freedom of the massive vector $A_\mu$ \cite{Heisenberg:2014rta,Tasinato:2014eka,Allys:2015sht,BeltranJimenez:2016rff,Allys:2016jaq}.
Only the following $G_2$-$G_4$ sector is needed for the rotating solution:
\begin{equation}\label{Action}
\begin{aligned}
\mathcal{S}=\int d^4x\sqrt{-g}\,\bigg\{
G_2(X)+G_3(X)\nabla_\mu A^\mu+G_4(X)R
+G_{4,X}(X)\left[
(\nabla_\mu A^\mu)^2-\nabla_\rho A_\sigma\nabla^\sigma A^\rho
\right]\bigg\},
\end{aligned}
\end{equation}
where $X\equiv-\frac{1}{2}A_\mu A^\mu$.
This restriction is not an ad hoc truncation introduced for the black-hole ansatz.
Four-dimensional Proca theories obtained by regularizing the Lovelock invariants fall precisely within Eq. \eqref{Action} \cite{Charmousis:2025proca,Fernandes:2025lovelockproca}.
Up to a common normalization, the $n$-th Lovelock density generates
\begin{equation}\label{LovelockProcaFunctions}
\begin{aligned}
G_2^{(n)}(X)=-6\,2^n(n-1)X^n,\qquad
G_3^{(n)}(X)=-2^{n+1}nX^{n-1},\qquad
G_4^{(n)}(X)=-\frac{2^{n-1}n}{2n-3}X^{n-1}.
\end{aligned}
\end{equation}
The Gauss-Bonnet contribution corresponds to $n=2$ in Eq. \eqref{LovelockProcaFunctions}, which also specifies the vector interactions inherited from each higher-order Lovelock density \cite{Charmousis:2025proca,Fernandes:2025lovelockproca}.
Thus, the Lovelock-Proca models form a theoretically distinguished subset of Generalized Proca theory, while the exact rotating construction applies to the slightly broader near-$X=0$ class
\begin{equation}\label{GPfunctions}
\begin{aligned}
G_2(X)=\mathcal{O}(X^2),\quad
G_3(X)=\alpha c_{3,1}X+\mathcal{O}(X^2),\quad
G_4(X)=1+\alpha c_{4,1}X+\alpha^2c_{4,2}X^2
+\mathcal{O}(X^3).
\end{aligned}
\end{equation}
Here $\alpha$ has dimensions of length squared, whereas $c_{3,1}$, $c_{4,1}$, and $c_{4,2}$ are dimensionless couplings.
This near $X=0$ expansion is sufficient for the rotating construction because the Proca configuration is null on shell \cite{Fernandes:2026primary}.

In the exact solutions of Ref. \cite{Fernandes:2026primary}, the metric and Proca field share a null direction: the geometry has Kerr-Schild form, and the vector field lies along its null congruence.
This alignment also appears in the Kerr-Newman construction \cite{Newman:1965my}.
In ingoing Kerr coordinates $(v,r,\theta,\tilde{\varphi})$, the line element is
\begin{equation}\label{primaryKerrSchild}
\begin{aligned}
ds^2={}&\!-\!\Big[1\!-\!\frac{2rM(r,\theta)}{\Sigma }\Big]
(dv+a\sin^2\theta\,d\tilde{\varphi})^2
+\Sigma\sin^2\theta\,d\tilde{\varphi}^{\,2}
+\Sigma\, d\theta^2\\
&+2(dv+a\sin^2\theta\,d\tilde{\varphi})
(dr+a\sin^2\theta\,d\tilde{\varphi}).
\end{aligned}
\end{equation}
Here, $\Sigma=r^2+a^2\cos^2\theta$ and $a$ is the rotation parameter.
The corresponding vector one-form is
\begin{equation}
A_\mu dx^\mu=
\frac{\mathcal{A}(r,\theta)}{\Sigma}
\left(dv+a\sin^2\theta\,d\tilde{\varphi}\right).
\end{equation}
This configuration is null on shell, $X=0$, and reduces the independent field equations to a first-order system whose solution is encoded in the generalized mass function $M(r,\theta)$.
The metric and the aligned Proca field therefore constitute an exact solution of the underlying vector-tensor field equations, rather than a phenomenological deformation introduced at the metric level.
Within Eq. \eqref{primaryKerrSchild}, the distinction between circular and noncircular geometries is controlled by the angular dependence of $M$: circularity requires $\partial_\theta M=0$ \cite{Fernandes:2026primary,Babichev:2025szb}.

\subsection{Primary-hair geometry considered in this work}

The general Kerr-Schild construction above admits several analytic branches.
We focus on the $c_{3,1}=0$ branch of Ref. \cite{Fernandes:2026primary}, for which the radial Proca equation can be integrated analytically.
Writing $\chi=\cos\theta$, the resulting Proca profile is
\begin{equation}
\mathcal{A}(r,\chi)=q(\chi)^{-1/2}
r^{\frac{c_{4,1}^2-2c_{4,2}}
{2(c_{4,1}^2+2c_{4,2})}}
\Sigma^{\frac{c_{4,1}^2+4c_{4,2}}
{2(c_{4,1}^2+2c_{4,2})}},
\end{equation}
where $q(\chi)$ is an arbitrary integration function reflecting the integration freedom in the Proca sector.
Requiring the ADM mass to be independent of the polar angle reduces it to an arbitrary constant, $q(\chi)=q$.
Since $q$ is not fixed by the ADM mass or angular momentum, while still modifying the geometry through the generalized mass function, it constitutes primary hair \cite{Heisenberg:2017xda,Charmousis:2026dbi}.
The corresponding generalized mass function is given by
\begin{equation}\label{generalMassFunction}
M(r,\theta)=m+\frac{\alpha c_{4,1}}{2q}
\left[
\left(1+\frac{a^2\cos^2\theta}{r^2}\right)^{
\frac{2c_{4,2}}{c_{4,1}^2+2c_{4,2}}}
-1
\right],
\end{equation}
where $m$ is the ADM mass and $q$ has dimensions of length.
Together with the Proca profile given above, Eqs. \eqref{primaryKerrSchild} and \eqref{generalMassFunction} specify the $c_{3,1}=0$ branch of the exact solution presented in Eqs. (3.1), (4.7), and (4.10) of Ref. \cite{Fernandes:2026primary}.

To isolate a one-parameter deformation of Kerr, we assume $q\neq0$ and specialize to the nondegenerate coupling slice $ c_{4,2}=-c_{4,1}^2 $, with $c_{4,1}\neq0$.
On this slice, the exponent $2c_{4,2}/(c_{4,1}^2+2c_{4,2})$ in Eq. \eqref{generalMassFunction} equals two, and Ref. \cite{Fernandes:2026primary} explicitly exhibited an angle-dependent horizon for a representative parameter choice.
The constants that enter the metric can then be collected into the dimensionless quantity
\begin{equation}\label{hairparameter}
\xi\equiv\frac{\alpha c_{4,1}}{2qm},
\end{equation}
so that
\begin{equation}\label{simpleMassFunction}
M(r,\theta)
=m\left\{
1+\xi\left[
\left(1+\frac{a^2\cos^2\theta}{r^2}\right)^2-1
\right]\right\}.
\end{equation}
The geometry is now governed by $(m,a,\xi)$, with $\xi$ measuring the effective strength of the primary-hair deformation on the chosen theory slice.
This parametrization has two useful limits.
First, $\xi=0$ gives $M(r,\theta)=m$ and restores Kerr exactly \cite{Kerr:1963ud}.
Second, the correction vanishes for $a=0$ even at nonzero $\xi$; the metric is then Schwarzschild although the Proca field can remain nontrivial.
The departure from the general-relativistic geometry is therefore genuinely rotation activated.

To express this solution in a form that reduces directly to the standard Boyer-Lindquist representation of Kerr \cite{Boyer:1966qh}, define
\begin{equation}
\Delta_{\rm Kerr}=r^2-2mr+a^2,
\qquad
\delta M(r,\theta)=M(r,\theta)-m,
\end{equation}
and apply the Kerr coordinate transformation
\begin{equation}\label{KerrCoordinateTransformation}
dv=dt+\frac{r^2+a^2}{\Delta_{\rm Kerr}}\,dr,
\qquad
d\tilde{\varphi}=-d\varphi-\frac{a}{\Delta_{\rm Kerr}}\,dr .
\end{equation}
The sign of $\varphi$ has been chosen so that the Kerr limit follows the usual Boyer-Lindquist convention.
The Kerr part of the transformed metric is
\begin{equation}\label{KerrBL}
\begin{aligned}
ds^2_{\rm Kerr}={}-\frac{\Delta_{\rm Kerr}}{\Sigma}
\left(dt-a\sin^2\theta\,d\varphi\right)^2
+\frac{\Sigma}{\Delta_{\rm Kerr}}\,dr^2
+\Sigma\,d\theta^2
+\frac{\sin^2\theta}{\Sigma}
\left[a\,dt-(r^2+a^2)d\varphi\right]^2 .
\end{aligned}
\end{equation}
Combining Eqs. \eqref{simpleMassFunction}-\eqref{KerrBL}, the selected primary-hair solution takes the compact form
\begin{equation}\label{ds2}
\begin{aligned}
ds^2=ds_{\rm Kerr}^2
+\frac{2r\,\delta M}{\Sigma}
\left(
dt-a\sin^2\theta\,d\varphi+\frac{\Sigma}{\Delta_{\rm Kerr}}\,dr
\right)^2 .
\end{aligned}
\end{equation}
Equation \eqref{ds2} is the desired Kerr-adapted representation of the exact primary-hair geometry.
For $\xi=0$, $\delta M=0$ and the correction vanishes identically.
For $\xi\neq0$, the explicit angular dependence of $M$ makes the spacetime noncircular, so Eq. \eqref{ds2} should be regarded as Boyer-Lindquist-like rather than as a genuine Boyer-Lindquist form \cite{Babichev:2025szb}.
For later use, its nonvanishing covariant components are, with the arguments of $M(r,\theta)$ and $\delta M(r,\theta)$ suppressed,
\begin{equation}\label{primaryMetricComponents}
\begin{aligned}
&g_{tt}=-1\!+\!\frac{2rM}{\Sigma}, \quad
g_{tr}=\frac{2r\,\delta M}{\Delta_{\rm Kerr}},\quad
g_{t\varphi}=\!-\!\frac{2arM\sin^2\theta}{\Sigma}, 
\\
&g_{rr}=\frac{\Sigma}{\Delta_{\rm Kerr}}
+\frac{2r\Sigma\,\delta M}{\Delta_{\rm Kerr}^2},\quad
g_{r\varphi}=-\frac{2ar\,\delta M\sin^2\theta}{\Delta_{\rm Kerr}},
\\
&g_{\theta\theta}=\Sigma,\quad 
g_{\varphi\varphi}=
\Big(r^2+a^2+\frac{2a^2rM\sin^2\theta}{\Sigma} \Big)\sin^2\theta.
\end{aligned}
\end{equation}
Thus the primary hair not only modifies the Kerr components but also activates $g_{tr}$ and $g_{r\varphi}$ in these coordinates, both of which vanish in the Kerr limit.

\begin{figure}[h]
\centering
\includegraphics[width=0.45\linewidth]{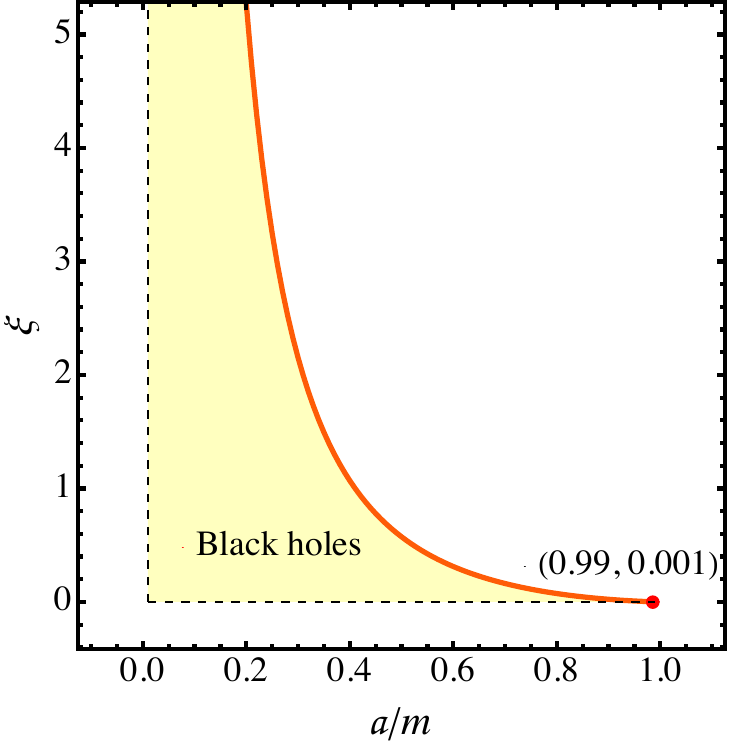}
\caption{Numerical horizon-admitting region in the $(a/m,\xi)$ plane obtained from Eq. \eqref{deformedHorizon} using the fourth-order Runge-Kutta method.
The shaded area represents smooth outer-horizon solutions, and the orange curve marks their critical boundary.}
\label{fig1}
\end{figure}

The factors $\Delta_{\rm Kerr}^{-1}$ in Eqs. \eqref{ds2} and \eqref{primaryMetricComponents} are inherited from the coordinate transformation \eqref{KerrCoordinateTransformation}; consequently, $\Delta_{\rm Kerr}=0$ is a coordinate singularity of the Kerr-adapted form and does not, for $\xi\neq0$, locate the physical event horizon.
Because the spacetime is noncircular, its horizon need not lie at a constant radial coordinate \cite{Babichev:2025szb,DelPorro:2025hse}.
We therefore parameterize a stationary and axisymmetric candidate horizon in the regular ingoing coordinates by $F(r,\theta)=r-H(\theta)=0$.
Imposing the null-surface condition $g^{\mu\nu}\partial_\mu F\,\partial_\nu F=0$ and using $g^{rr}=\Delta/\Sigma$, $g^{r\theta}=0$, and $g^{\theta\theta}=1/\Sigma$ gives
\begin{equation}\label{deformedHorizon}
\begin{aligned}
\Big[\frac{dH(\theta)}{d\theta}\Big]^2+
\Delta\bigl(H(\theta),\theta\bigr)=0,\qquad
\Delta(r,\theta)\equiv r^2+a^2-2rM(r,\theta).
\end{aligned}
\end{equation}
In the numerical construction that follows, we restrict to $\xi\geq0$; all subsequent calculations are therefore performed within the horizon-admitting region of this parameter domain.
For the reflection-symmetric solution considered here, the outer horizon can be obtained on $0\leq\theta\leq\pi/2$ with $H'(0)=H'(\pi/2)=0$ and extended to the other hemisphere by symmetry.
The outermost regular solution is identified with the outer event horizon.
For each point in the $(a/m,\xi)$ plane, we integrate Eq. \eqref{deformedHorizon} numerically with a fourth-order Runge-Kutta method to determine whether a regular angle-dependent horizon exists and, when it does, to construct its profile $H(\theta)$.
A parameter point represents a black hole when the integration yields a real, smooth outer solution that extends over the entire interval $0\leq\theta\leq\pi/2$ and satisfies the equatorial regularity condition.
Repeating this calculation across the parameter plane therefore determines both the horizon profiles and their domain of existence.
Details of the numerical implementation are given in Appendix \ref{app:horizon_numerics}.
The admissible domain in the $(a/m,\xi)$ plane is thus determined by regular outer-horizon solutions spanning the full angular interval.
Since Eq. \eqref{deformedHorizon} is also invariant under $a\rightarrow-a$, it is sufficient to display the nonnegative-spin domain; the same horizon bound applies to negative values of $a$.
The resulting horizon-admitting domain in the nonnegative-spin sector is shown in Fig. \ref{fig1}.

For efficient parameter selection in the subsequent calculations, we approximate the numerically determined boundary over $0<a/m\leq0.99$ with $\xi\geq 0$ by a Chebyshev fit.
Defining $x\equiv\sqrt{1-(a/m)^2}$, the resulting practical criterion is
\begin{equation}\label{admissibleParameterRegion}
\begin{aligned}
\left[(a/m)^{-2}-1\right]
\bigl(0.03881834+0.10126918x+0.09799828x^2-0.01352359x^3\bigr)-\xi>0.
\end{aligned}
\end{equation}
As $a/m\rightarrow0$, its upper boundary diverges, while the $\xi$-dependent deformation vanishes and the geometry approaches Schwarzschild for any fixed $\xi$.

In this work, we perform the ray tracing directly in the Kerr-adapted coordinates using the metric \eqref{ds2}.
Equation \eqref{deformedHorizon} determines the angle-dependent capture surface $r=H(\theta)$, which is used to identify photons falling into the black hole.
The coordinate singularity at $\Delta_{\rm Kerr}=0$ is monitored separately and is not used as a horizon condition.
Stationarity and axial symmetry still guarantee conservation of the photon energy and axial angular momentum.
Since the explicit $\theta$ dependence of $M(r,\theta)$ generically obstructs the Kerr Hamilton-Jacobi separation, the photon trajectories are integrated numerically \cite{Babichev:2025szb}.

\section{Forward modeling of black hole shadows}\label{sec:forward}

\subsection{Ray tracing method}

We construct the shadow by tracing null geodesics backward from a distant observer through the primary-hair geometry \eqref{ds2}.
In Hamiltonian form, the photon equations follow from
\begin{equation}\label{EqHJ}
\mathcal{H}(x,p)=\frac{1}{2}g^{\mu\nu}p_\mu p_\nu=0,
\quad
\dot{x}^{\mu}=\frac{\partial\mathcal{H}}{\partial p_\mu},
\quad
\dot{p}_{\mu}=-\frac{\partial\mathcal{H}}{\partial x^\mu},
\end{equation}
where an overdot denotes differentiation with respect to an affine parameter.
Stationarity and axial symmetry give the two conserved momenta
\begin{equation}\label{gabE}
p_t=-\mathcal{E},
\qquad
p_\varphi=L .
\end{equation}
Writing $p_\mu=(-\mathcal{E},p_r,p_\theta,L)$, the null Hamiltonian can be expanded as
\begin{equation}
\begin{aligned}
\mathcal{H}={}
\frac{1}{2}g^{tt}\mathcal{E}^2
-g^{tr}\mathcal{E}p_r
-g^{t\varphi}\mathcal{E}L
+\frac{1}{2}g^{rr}p_r^2
+g^{r\varphi}p_rL
+\frac{1}{2}g^{\theta\theta}p_\theta^2
+\frac{1}{2}g^{\varphi\varphi}L^2=0.
\end{aligned}
\end{equation}
Hamilton's equations then give the remaining first-order system
\begin{equation}\label{tauvarphi}
\begin{aligned}
\dot{t}&=-g^{tt}\mathcal{E}+g^{tr}p_r+g^{t\varphi}L,\qquad
\dot{r}=-g^{tr}\mathcal{E}+g^{rr}p_r+g^{r\varphi}L,\qquad
\dot{\theta}=g^{\theta\theta}p_\theta,\\
\dot{\varphi}&=-g^{t\varphi}\mathcal{E}
+g^{r\varphi}p_r+g^{\varphi\varphi}L,\qquad
\dot{p}_r=-\frac{1}{2}\,\partial_r g^{\mu\nu}p_\mu p_\nu,\qquad
\dot{p}_\theta=-\frac{1}{2}\,\partial_\theta g^{\mu\nu}p_\mu p_\nu .
\end{aligned}
\end{equation}
The inverse components in Eq. \eqref{tauvarphi} are obtained from the covariant metric \eqref{primaryMetricComponents}.
For the present geometry, $g^{r\theta}=0$, whereas $g^{tr}$ and $g^{r\varphi}$ are generally nonzero.
These components couple the radial motion to the temporal and azimuthal equations.
Moreover, the explicit $(r,\theta)$ dependence of the metric generally prevents a Carter-type separation, so Eq. \eqref{tauvarphi} is integrated numerically.

We use backward ray tracing, with technical details described in Refs. \cite{Wang:2017qhh,Hu:2020usx,Zhong:2021mty,Garnier:2023lph}.
Each pixel on the observer's screen fixes a local photon direction, which is converted into coordinate momentum and evolved with Eq. \eqref{tauvarphi}.
Here ``backward'' refers to reconstructing the source from the image by launching computational rays from the observer into the scene.
The computational rays are past-directed and are integrated from the observer with an increasing affine parameter; accordingly, their local temporal component satisfies $p^{(t)}<0$.
To define this local frame, consider the spatial metric on a $t=\mathrm{const}$ slice and introduce
\begin{equation}
\begin{aligned}
&\mathcal{D}\equiv g_{rr}g_{\varphi\varphi}-g_{r\varphi}^{\,2},\qquad\,\,\,
\beta^r\equiv
\frac{g_{\varphi\varphi}g_{tr}-g_{r\varphi}g_{t\varphi}}
{\mathcal{D}},  \\
&\beta^\varphi\equiv
\frac{g_{rr}g_{t\varphi}-g_{r\varphi}g_{tr}}
{\mathcal{D}},\quad
\mathcal{N}^2\equiv
-g_{tt}+\beta^r g_{tr}+\beta^\varphi g_{t\varphi}.
\end{aligned}
\end{equation}
An orthonormal tetrad adapted to the Eulerian observer is then
\begin{equation}\label{zjbj}
\begin{aligned}
e_{(t)}&=\frac{1}{\mathcal{N}}
\left(\partial_t-\beta^r\partial_r-\beta^\varphi\partial_\varphi\right),  \qquad e_{(\theta)}=\frac{1}{\sqrt{g_{\theta\theta}}}\partial_\theta\\
e_{(r)}&=-\sqrt{\frac{g_{\varphi\varphi}}{\mathcal{D}}}
\left(\partial_r-\frac{g_{r\varphi}}{g_{\varphi\varphi}}
\partial_\varphi\right),\qquad
e_{(\varphi)}=\frac{1}{\sqrt{g_{\varphi\varphi}}}\partial_\varphi ,
\end{aligned}
\end{equation}
where all metric components are evaluated at the observer position $(r_0,\theta_0)$.
The minus sign in $e_{(r)}$ makes the radial basis vector point toward the black hole.
This tetrad satisfies $g_{\mu\nu}e_{(a)}^{\mu}e_{(b)}^{\nu}=\eta_{(a)(b)}$ and reduces to the usual Kerr--Boyer--Lindquist Eulerian tetrad when $\xi=0$.

The local tetrad components of the computational ray are $p^{(t)}=-p_\mu e_{(t)}^\mu$ and $p^{(i)}=p_\mu e_{(i)}^\mu$, with $i=r,\theta,\varphi$.
Conversely,
\begin{equation*}
p^\mu=p^{(t)}e_{(t)}^\mu+p^{(r)}e_{(r)}^\mu
+p^{(\theta)}e_{(\theta)}^\mu+p^{(\varphi)}e_{(\varphi)}^\mu .
\end{equation*}
The canonical momenta entering the Hamiltonian system are then obtained from $p_\mu=g_{\mu\nu}p^\nu$ using the full metric.
This operation includes the contributions from $g_{tr}$ and $g_{r\varphi}$ and yields the conserved quantities defined in Eq. \eqref{gabE}.
With the past-directed convention adopted here, $\mathcal{E}<0$; the corresponding physical future-directed photon has positive energy $\mathcal{E}_{\rm phys}=-\mathcal{E}>0$.
For a past-directed null ray, $p^{(t)}=-|\boldsymbol{P}|$, while its spatial direction is parameterized by
\begin{equation}\label{obsangles}
\begin{aligned}
p^{(r)}=|\boldsymbol{P}|\cos\alpha\cos\beta,\qquad
p^{(\theta)}=|\boldsymbol{P}|\sin\alpha,\qquad
p^{(\varphi)}=|\boldsymbol{P}|\cos\alpha\sin\beta .
\end{aligned}
\end{equation}
The overall momentum scale is irrelevant for null trajectories and may be set to unity.
The associated screen coordinates are
\begin{equation}\label{screenxy}
\begin{aligned}
X=-r_0\tan\beta
=-r_0\frac{p^{(\varphi)}}{p^{(r)}},\qquad
Y=r_0\frac{\tan\alpha}{\cos\beta}
=r_0\frac{p^{(\theta)}}{p^{(r)}} .
\end{aligned}
\end{equation}
The geodesic equations are integrated with an increasing affine parameter using the past-directed initial four-momentum constructed above.
A ray is classified as captured when it reaches the angle-dependent horizon $r=H(\theta)$ determined by Eq. \eqref{deformedHorizon}; otherwise it is continued until it intersects the outer celestial sphere.
The shadow boundary is formed by the critical rays separating these two outcomes \cite{Liu:2026optical,Cunha:2018acu,Perlick:2021aok}.

\begin{figure}[htbp]
\centering
\begin{minipage}[c]{0.58\linewidth}
\centering
\includegraphics[width=\linewidth]{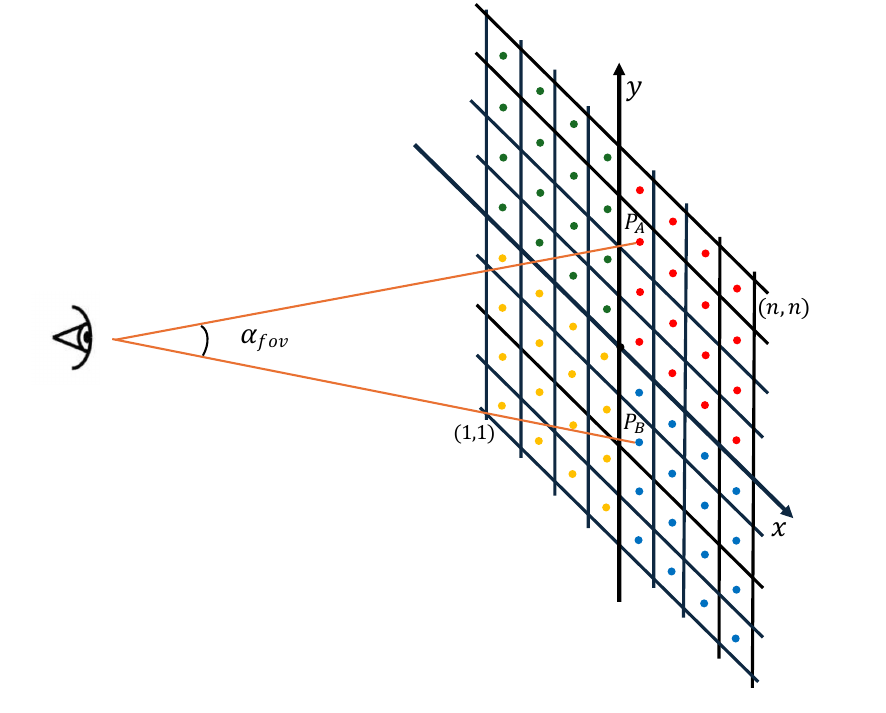}
\end{minipage}\hfill
\begin{minipage}[c]{0.38\linewidth}
\centering
\includegraphics[width=\linewidth]{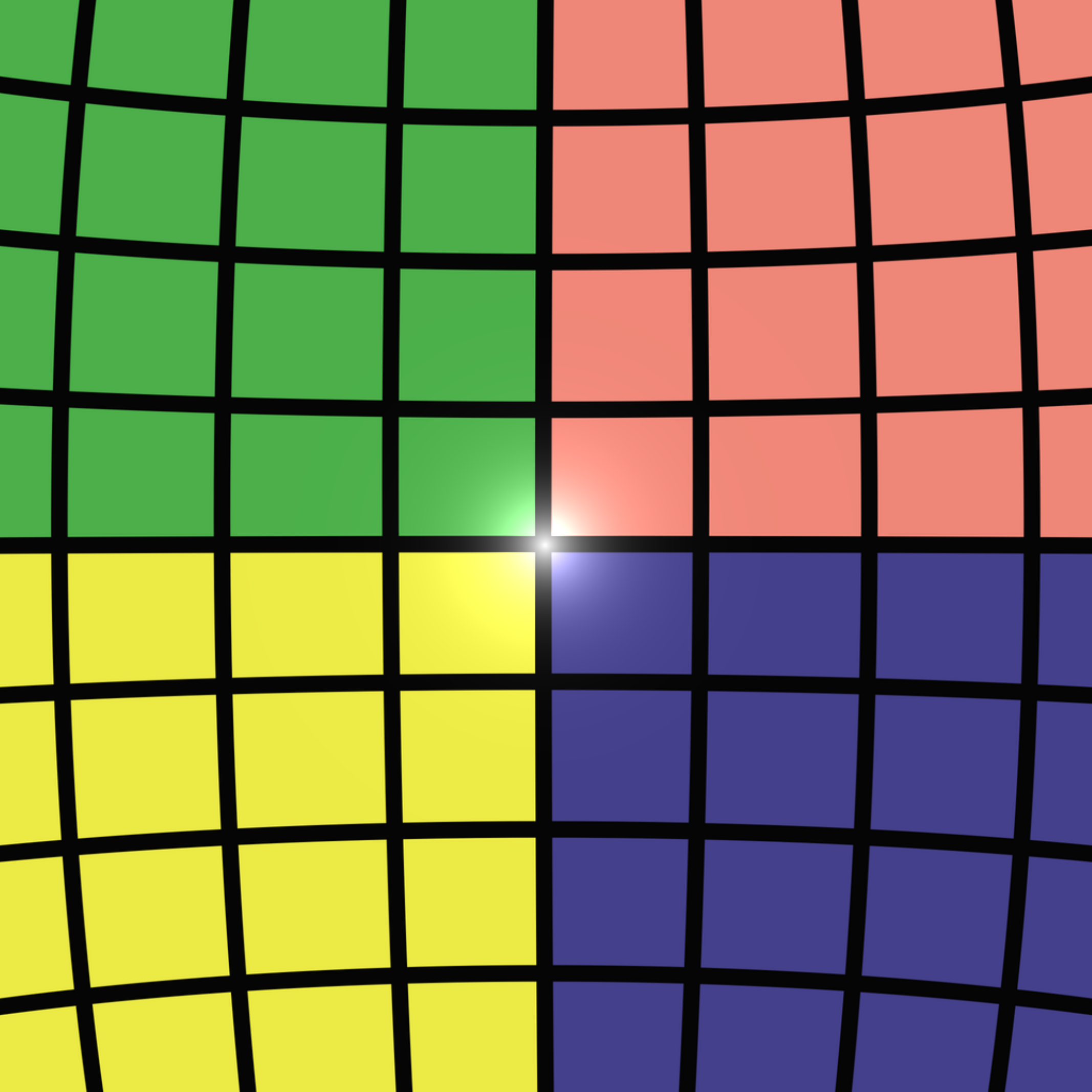}
\end{minipage}
\caption{Left: Observer-screen construction used for backward ray tracing.
The field of view $\alpha_{\rm fov}$ is sampled by an $n\times n$ grid on the local $(x,y)$ screen, with $P_A$ and $P_B$ illustrating two representative pixel centers.
Right: Synthetic background used on the celestial sphere, constructed following the four-color grid setup of Kuang et al. \cite{Kuang:2024ugn}.
The color boundaries and black grid provide reference markers for identifying lensing distortions and parity changes.}
\label{fig2}
\label{fig3}
\end{figure}

Figure \ref{fig2} (left) summarizes the initialization of the ray-tracing calculation.
An observer at $(r_0,\theta_0)$ views the spacetime through a local screen with field of view $\alpha_{\rm fov}$, discretized into $n\times n$ pixels.
Here $\theta_0$ is measured from the rotation axis and specifies the viewing inclination, whereas $(m,a,\xi)$ characterize the black-hole geometry.
Unless stated otherwise, we set $\theta_0=\pi/2$.
For each pixel center, Eqs. \eqref{screenxy} and \eqref{obsangles} determine the local photon direction, which is converted into coordinate initial data with the tetrad \eqref{zjbj}.
Rays intercepted by the horizon form the black shadow.
Escaping rays are continued to the celestial sphere and assigned the color of their intersection point on the synthetic background in Fig. \ref{fig3} (right).
Following Ref. \cite{Kuang:2024ugn}, the black hole and observer are placed inside this distant spherical source, which is divided into four colored regions with a white marker at their common intersection and a black longitude--latitude reference grid.
These markers make image distortion, multiple imaging, and parity changes readily visible.

\subsection{Black hole shadow and gravitational lensing}

\begin{figure*}[t]
\centering
\includegraphics[width=0.9\textwidth]{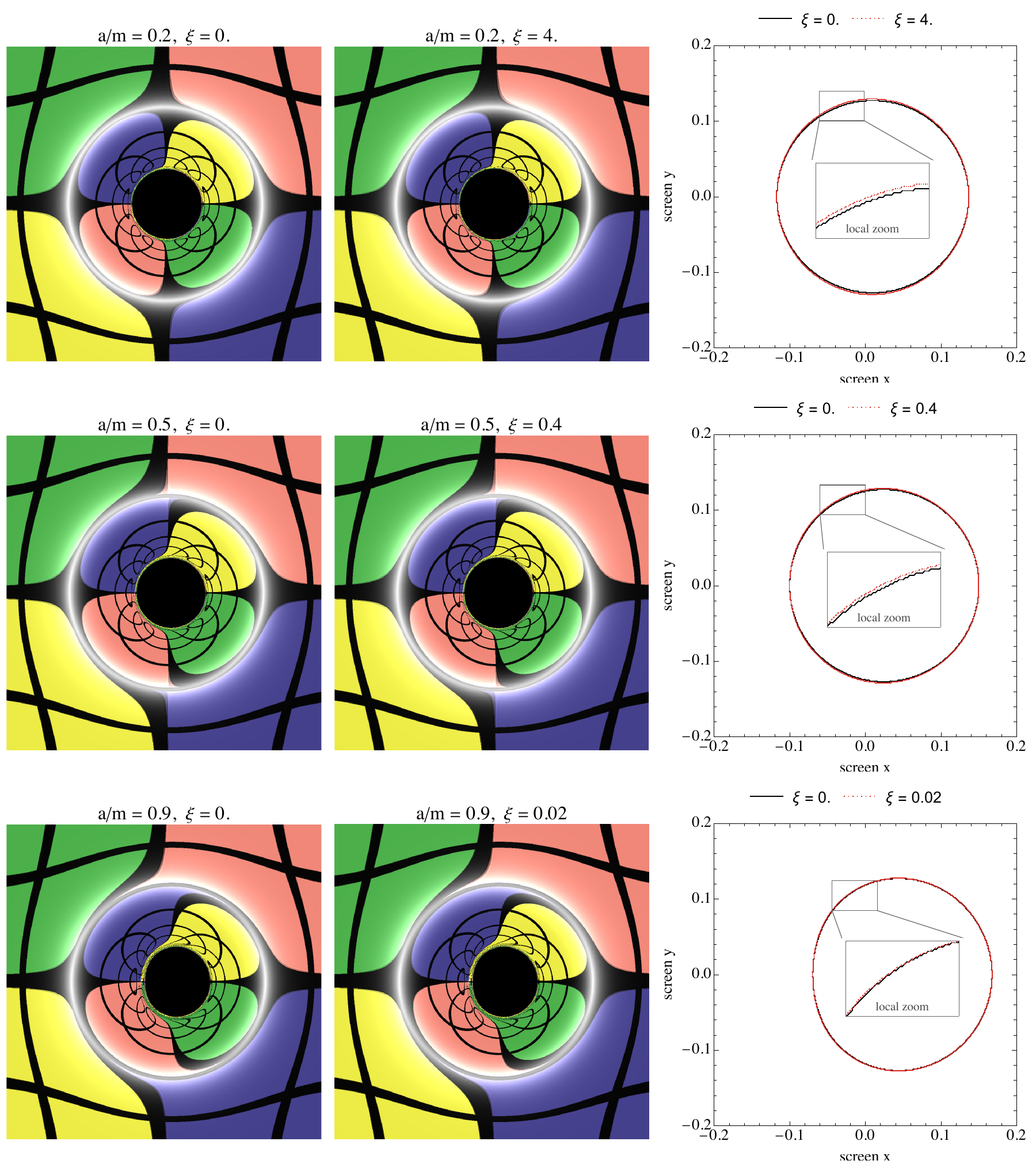}
\caption{{Ray-traced shadow and lensing images} ($1024\times1024$ rays) using the background in Fig.~\ref{fig3} (right), with $r_0=40m$, $\theta_0=\pi/2$, and $\alpha_{\rm fov}=\pi/3$.
Rows correspond to $a/m=0.2$, $0.5$, and $0.9$.
{Columns show Kerr, the selected primary-hair configuration, and their overlaid capture boundaries.} {Insets show equal-sized upper-half windows with maximal changed-pixel counts in the primary-hair minus Kerr masks.}}
\label{fig4}
\end{figure*}

{
Figure~\ref{fig4} illustrates the black hole shadows and gravitational lensing patterns for an equatorial observer at $r_0=40m$, with a field of view $\alpha_{\rm fov}=\pi/3$.
Each row compares Kerr with a selected primary-hair configuration at $a/m=0.2$, $0.5$, or $0.9$.
Following the Kerr column isolates the effects of rotation.
As the spin increases, the shadow becomes progressively displaced toward positive screen $x$, and its left edge becomes flatter, producing a more pronounced D-shaped outline at $a/m=0.9$.
Different behavior of co-rotating and counter-rotating photon trajectories underlies this horizontal asymmetry, while equatorial reflection symmetry preserves the symmetry of the upper and lower boundaries \cite{Perlick:2021aok}.
\par}

{
At each fixed spin, the primary-hair image retains the overall morphology of its Kerr counterpart, while the overlaid contours reveal a small enlargement of the captured region.
Along the upper-left segments shown in the enlarged views, the primary-hair contour lies slightly outside the Kerr contour, with the clearest separation in the first two rows.
Expanding the mass function in Eq.~\eqref{simpleMassFunction} gives:
\begin{equation*}
\frac{\delta M}{m}
=2\xi\frac{a^2\cos^2\theta}{r^2}
+\xi\frac{a^4\cos^4\theta}{r^4}.
\end{equation*}
Rotation thus activates a latitude-dependent correction that grows toward smaller radii at fixed polar angle.
Through the metric components, this correction changes the radial and azimuthal motion of photons and introduces additional angular dependence into their propagation.
Although the observer is located on the equatorial plane, rays reaching pixels away from the horizontal screen axis generally travel through regions with $\theta\ne\pi/2$, where the correction is nonzero.
For the selected parameter pairs, the leading coefficient $\xi(a/m)^2$ takes the values $0.16$, $0.10$, and $0.016$ respectively, setting the leading mass-function correction at a common $r/m$ and $\theta$.
\par}

{
Outside the shadow, gravitational lensing bends the background grid into extended arcs.
Repeated, increasingly crowded structures near the boundary display the multiple imaging produced by strongly deflected rays \cite{Gralla:2019xty}.
Increasing the spin makes this pattern more asymmetric: the central dark region shifts relative to the broad colored structures, and the grid arcs become distributed unevenly on its two sides.
Here the source-to-screen mapping depends on the direction of photon motion relative to the black hole rotation.
At fixed spin, the primary-hair panels retain the main lensing pattern, with small adjustments of the grid arcs near the shadow.
Because $\delta M$ depends on both radius and latitude, rays passing close to the black hole through off-equatorial regions are particularly sensitive to the hair correction.
Escaping rays accumulate changes in direction along these paths before reaching the source sphere.
Together, the capture boundary and distorted background display how the deformed geometry affects both captured and escaping rays.
\par}

{
To quantify the overall size changes, we compare the captured-pixel counts.
For images generated with identical observer and screen settings, let $N_\xi$ and $N_0$ denote the counts for the primary-hair and Kerr configurations at the same spin.
We define
\begin{equation*}
\delta_A=\frac{N_\xi}{N_0}-1,
\qquad
\delta_R=\sqrt{\frac{N_\xi}{N_0}}-1.
\end{equation*}
With the common pixel-area factor canceled, these ratios measure the changes relative to Kerr.
As listed in Table~\ref{tab:shadow_preview}, the captured area increases by approximately $1.8\%$, $1.2\%$, and $0.3\%$ for the three parameter pairs, accompanied by area-equivalent radius increases of approximately $0.9\%$, $0.6\%$, and $0.2\%$.
For such small changes, $\delta_R\simeq\delta_A/2$, explaining the smaller fractional response of the equivalent radius.
A modest net area increase therefore accompanies the local boundary and lensing changes visible in Fig.~\ref{fig4}.
\par}

\begin{table}[H]
\centering
\renewcommand{\arraystretch}{1.2}
\begin{tabular*}{0.70\linewidth}{@{\extracolsep{\fill}}cccc@{}}
\hline\hline
$a/m$ & Compared $\xi$ & Area change & Radius change\\
\hline
0.2 & $0\to4$ & $\approx +1.8\%$ & $\approx +0.9\%$\\
0.5 & $0\to0.4$ & $\approx +1.2\%$ & $\approx +0.6\%$\\
0.9 & $0\to0.02$ & $\approx +0.3\%$ & $\approx +0.2\%$\\
\hline\hline
\end{tabular*}
\caption{{Changes in captured area and area-equivalent radius relative to Kerr, from the $1024\times1024$ images in Fig.~\ref{fig4}.}}
\label{tab:shadow_preview}
\end{table}

\subsection{Shadow observables and forward map}\label{subsec:observables}
\begin{figure}[H]
\centering
\includegraphics[width=0.4\linewidth]{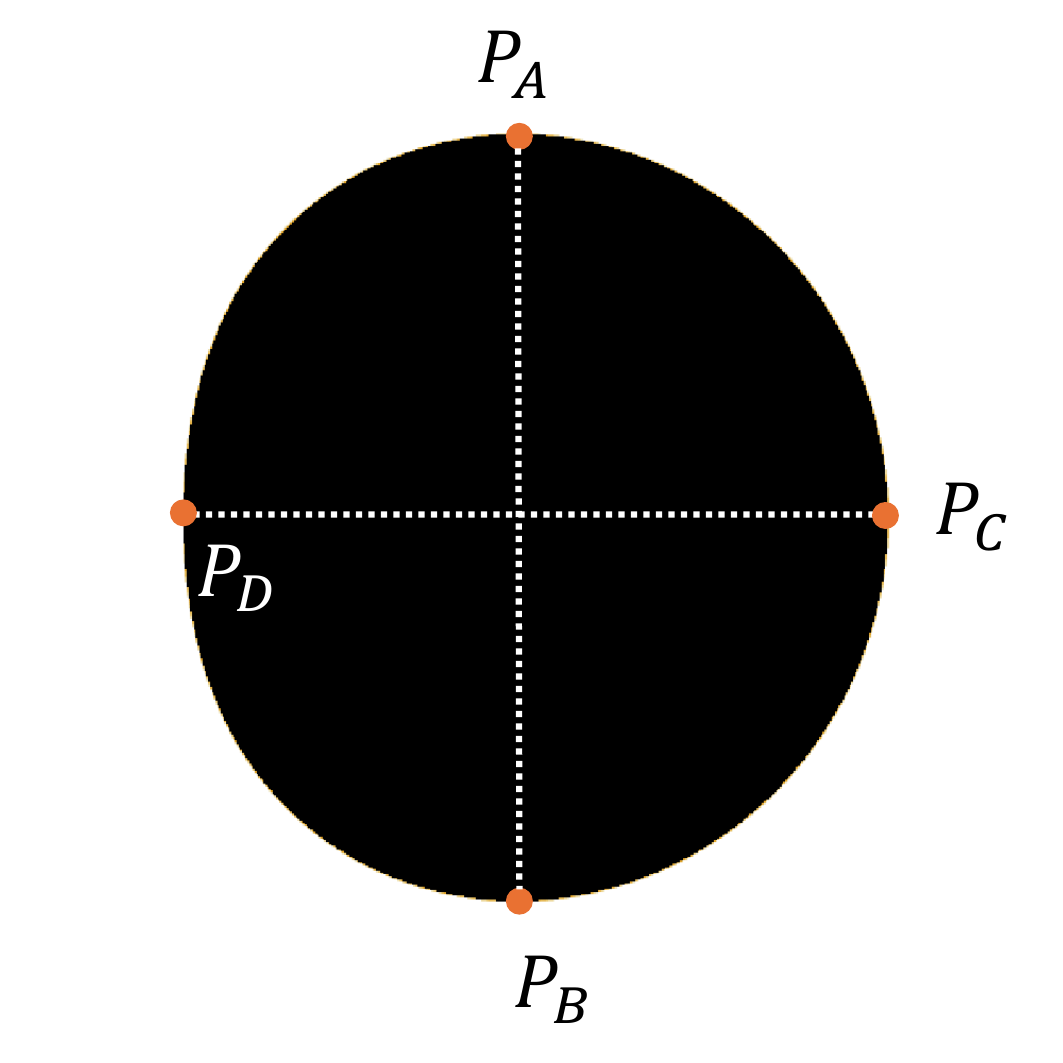}
\caption{Shadow area and characteristic diameters used to define the axis ratio.}
\label{fig5}
\end{figure}

{Building on the relative size comparison above, we characterize the shadows by their calibrated area and axis ratio for the subsequent forward modeling.} The shadow area $A_{\rm sh}$ measures the overall size of the captured region, while the axis ratio $\mathcal{D}_{\rm ax}$ retains its leading shape information.
Both are obtained directly from the image, without assuming an analytic form for the shadow boundary.
Their geometrical construction is shown in Fig. \ref{fig5}: the filled region determines the area, and its four extremal points define the vertical and horizontal diameters.

We evaluate the area by counting captured pixels.
Since the raw pixel count depends on the screen setup, each image is calibrated against a Schwarzschild shadow generated with the same observer position, field of view, and resolution.
If $N_{\rm sh}$ and $N_{\rm Sch}$ are the two captured-pixel counts, respectively, then
\begin{equation}\label{shadowArea}
 A_{\rm sh}=\frac{N_{\rm sh}}{N_{\rm Sch}}A_{\rm Sch},
 \qquad A_{\rm Sch}=27\pi m^2 .
\end{equation}
This internal calibration removes the pixel scale and cancels much of the discretization error shared by the two images.
It also uses the entire shadow mask, making $A_{\rm sh}$ less sensitive to the displacement of individual edge pixels than an estimator constructed only from boundary points.

The remaining shape information is extracted from the extremal points shown in Fig. \ref{fig5}.
Let $d_{(i,j)}$ be the Euclidean distance between two screen points.
The vertical and horizontal diameters are
\begin{equation}
 d_{\rm v}=d_{(A,B)},
 \qquad
 d_{\rm h}=d_{(C,D)},
\end{equation}
and we define
\begin{equation}\label{axisRatio}
 \mathcal{D}_{\rm ax}
 =\frac{\max\{d_{\rm v},d_{\rm h}\}}
        {\min\{d_{\rm v},d_{\rm h}\}} \geq 1 .
\end{equation}
The use of the larger diameter in the numerator makes the definition independent of the orientation of the elongation.
A circular shadow has $\mathcal{D}_{\rm ax}=1$, with departures from unity measuring its projected distortion.

Because $A_{\rm sh}$ scales as $m^2$, it is convenient to express the shadow size through the dimensionless area-equivalent radius
\begin{equation}\label{areaEquivalentRadius}
 \mathcal{R}_A
 \equiv\frac{1}{m}\sqrt{\frac{A_{\rm sh}}{\pi}}
 =\sqrt{27\,\frac{N_{\rm sh}}{N_{\rm Sch}}}.
\end{equation}
For a Schwarzschild black hole, $\mathcal{R}_A=3\sqrt{3}$ \cite{Perlick:2021aok}.
The pair $(\mathcal{R}_A,\mathcal{D}_{\rm ax})$ therefore separates the overall shadow scale from its leading distortion.

\begin{figure}[t]
\centering
\includegraphics[width=0.7\linewidth]{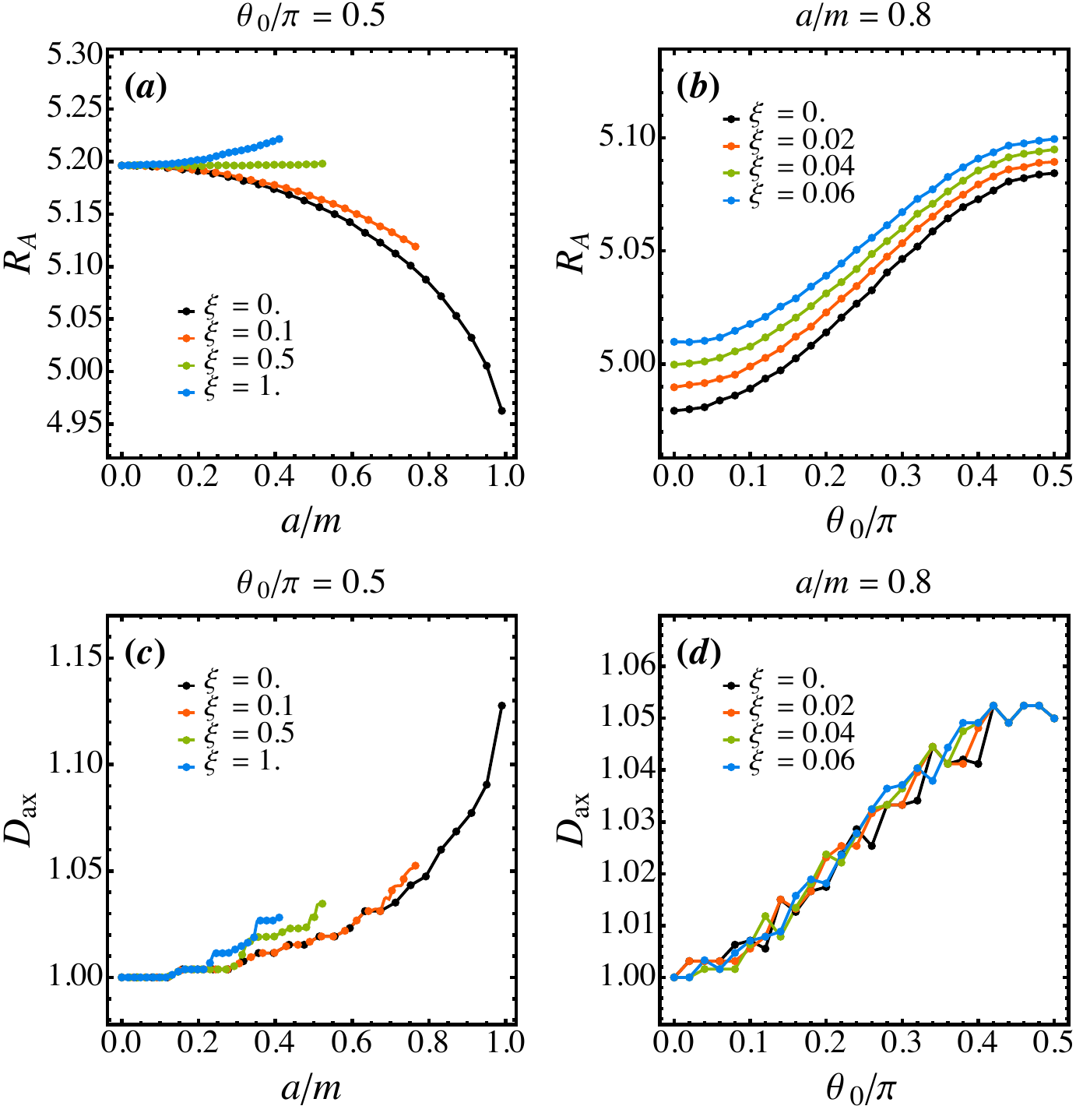}
\caption{{Area-equivalent radius $\mathcal{R}_A$ (top) and axis ratio $\mathcal{D}_{\rm ax}$ (bottom) as functions of spin for $\theta_0=\pi/2$ (left) and inclination for $a/m=0.8$ (right).
Colors indicate the fixed values of $\xi$ listed in each panel.
Spin sequences end at the boundary of the sampled parameter domain.}}
\label{fig6}
\end{figure}

{
The area-equivalent radius responds to both primary hair and viewing inclination, as shown in Figs.~\ref{fig6}(a) and \ref{fig6}(b).
For equatorial viewing, the Kerr radius decreases with increasing spin, whereas positive hair raises the radius and progressively weakens this decrease; an increasing trend appears for $\xi=1$ over the displayed range.
All curves meet at the Schwarzschild value in the nonrotating limit, consistent with the rotation-dependent metric correction.
At $a/m=0.8$, the radius increases overall toward equatorial viewing, while the separation between different hair configurations becomes smaller.
\par}

{
Figures~\ref{fig6}(c) and \ref{fig6}(d) show that the axis ratio increases overall with spin and inclination, reflecting the growing projected distortion of the shadow.
Axial viewing gives $\mathcal{D}_{\rm ax}=1$ for all displayed hair values, although the corresponding radii differ.
At $a/m=0.8$, the hair configurations are more clearly separated in radius than in axis ratio.
Size and shape therefore provide complementary information about the parameters, motivating their joint use in the subsequent Bayesian analysis.
\par}

Evaluating both quantities over the horizon-admitting domain of Sec. II yields the forward map
\begin{equation}\label{shadowForwardMap}
 (a/m,\xi,\theta_0)
 \longmapsto
 \left(\mathcal{R}_A,\mathcal{D}_{\rm ax}\right),
\end{equation}
which provides the input for the Bayesian analysis of Sec. \ref{sec3}.

\section{Bayesian constraints from M87* and Sgr A*}\label{sec3}

The ray-tracing calculations of Sec. III provide $\mathcal{R}_A$ and $\mathcal{D}_{\rm ax}$ throughout the allowed parameter space.
These two quantities form the link between the primary-hair geometry and the EHT measurements of M87* and Sgr A*.
For each object, we sample the parameter vector
\begin{equation}\label{inferenceParameters}
 \psi=\left\{a/m,\xi,\theta_0\right\}.
\end{equation}

The dimensionless radius $\mathcal{R}_A$ is converted into an angular scale using the mass $M_\bullet$ and distance $D_\bullet$ of the source.
The resulting theoretical angular diameter is
\begin{equation}\label{theoreticalAngularDiameter}
\begin{aligned}
 \vartheta_A^{\rm th}(\psi)
=2\mathcal{R}_A(\psi)\frac{G M_\bullet}{c^2D_\bullet}
=1.974196\times10^{-5}\,\mathcal{R}_A(\psi)
 \left(\frac{M_\bullet}{M_\odot}\right)
 \left(\frac{1\,{\rm kpc}}{D_\bullet}\right)\mu{\rm as}.
\end{aligned}
\end{equation}
At the observable level considered here, we identify this quantity with the measured EHT ring diameter, following the phenomenological prescription of Ref. \cite{Liu:2025wwq}, to provide a practical test of the shadow scale.

Table \ref{tab1} summarizes the observational inputs.
The M87* data consist of the latest ring-diameter measurement and the published upper limit on the axis ratio \cite{EventHorizonTelescope:2019ggy,EventHorizonTelescope:2025vum,EventHorizonTelescope:2019dse}.
For Sgr A*, we use the ring diameter and the mass--distance calibration adopted in Ref. \cite{Liu:2025wwq}, based on the 2022 EHT results \cite{EventHorizonTelescope:2022wkp}.
We fix $M_\bullet$ and $D_\bullet$ at their central values.
Their quoted uncertainties are retained in the table to indicate the external calibration accuracy, but are not propagated as nuisance parameters; the inferred constraints should consequently be understood as conditional on these central values.

\begin{table}[H]
\renewcommand{\arraystretch}{1.25}
\centering
\setlength\tabcolsep{1.8mm}
\begin{tabular}{lcc}
\hline\hline
Quantity & M87* & Sgr A*\\
\hline
Ring diameter $\vartheta^{\rm obs}$
& $43.9\pm0.6~\mu{\rm as}$
& $51.8\pm2.3~\mu{\rm as}$\\
Axis ratio $\mathcal{D}_{\rm ax}^{\rm obs}$
& $<4/3$ & ---\\
Mass $M_\bullet$
& $(6.5\pm0.7)\times10^9M_\odot$
& $(4.0\pm0.7)\times10^6M_\odot$\\
Distance $D_\bullet$
& $(16.8\pm0.8)~{\rm Mpc}$
& $(8.3\pm0.8)~{\rm kpc}$\\
\hline\hline
\end{tabular}
\caption{EHT observables and external calibration inputs adopted for the M87* and Sgr A* analyses.}
\label{tab1}
\end{table}

The physical domain established in Sec. II supplies the support of the prior.
Within the region covered by the numerical forward model, we adopt uniform priors
\begin{equation}\label{mcmcPriors}
 0\leq a/m\leq0.99,
 \qquad 0\leq\xi\leq5,
 \qquad 0\leq\theta_0\leq\frac{\pi}{2},
\end{equation}
and discard points that violate the horizon condition \eqref{admissibleParameterRegion}.
Here $\xi=5$ is the upper edge of the validated interpolation domain, whereas Eq. \eqref{admissibleParameterRegion} sets the actual physical boundary inside it.
{The joint prior is uniform within this admissible domain, but its one-dimensional marginals need not be uniform because the horizon condition couples spin and primary hair.
This restriction therefore contributes to the posterior structure for both sources.} Equatorial reflection symmetry allows us to restrict $\theta_0$ to one hemisphere.
We likewise sample only the nonnegative spin magnitude because the two observables are unchanged, within numerical accuracy, under the mirror transformation $a\rightarrow-a$.

With the theoretical predictions and observational inputs specified, we collect the angular-diameter and axis-ratio measurements in $d=\{\vartheta^{\rm obs},\sigma_{\vartheta}, \mathcal{D}_{\rm ax}^{\rm obs},\sigma_{\mathcal D}\}$ and write the effective likelihood, following Ref. \cite{Liu:2025wwq}, as
\begin{equation}\label{sourceLikelihood}
\begin{aligned}
 \mathcal{L}(d|\psi)={}&
 \mathcal{P}_{\vartheta}
 \bigl(\vartheta^{\rm obs},\sigma_{\vartheta}|\psi\bigr)
 \mathcal{P}_{\mathcal D}
 \bigl(\mathcal{D}_{\rm ax}^{\rm obs},
 \sigma_{\mathcal D}|\psi\bigr).
\end{aligned}
\end{equation}
For either observable $y\in\{\vartheta,\mathcal{D}_{\rm ax}\}$, a symmetric measurement uncertainty is represented by
\begin{equation}\label{observableGaussian}
 \mathcal{P}_{y}(y^{\rm obs},\sigma_y|\psi)
 \propto
 \exp\!\left[-\frac{
 \bigl(y^{\rm th}(\psi)-y^{\rm obs}\bigr)^2}
 {2\sigma_y^2}\right].
\end{equation}
One-sided bounds are represented by the corresponding one-sided probability, and an unavailable observable contributes no factor.
After substituting the appropriate entries from Table \ref{tab1}, the same construction is applied separately to M87* and Sgr A*.
Their likelihoods are not multiplied together, since $a/m$, $\xi$, and $\theta_0$ characterize the individual black hole.

Combining the likelihood with the prior gives the posterior
\begin{equation}\label{sourcePosterior}
 P(\psi|d)
 =\frac{\mathcal{L}(d|\psi)\Pi(\psi)}{\mathcal{Z}},
\end{equation}
where $\Pi$ is the prior in Eq. \eqref{mcmcPriors} and $\mathcal{Z}$ is the normalization.
Evaluating this expression by ray tracing every proposed parameter point would be prohibitively expensive \cite{Liu:2026latentimages,NaetheMotta:2025jipole}.
Instead, we first tabulate $\mathcal{R}_A$ and $\mathcal{D}_{\rm ax}$ across the admissible domain and interpolate the forward map \eqref{shadowForwardMap} during sampling.
Because this map is dimensionless, the same interpolation table applies to both sources; their mass--distance conversions and observational inputs are inserted only when evaluating the likelihood.
All three components of $\psi$ are varied simultaneously, and chain convergence is monitored with the Gelman--Rubin statistic \cite{Gelman:1992multiple}.
We retain only chains satisfying $R-1<0.02$.

\begin{figure}[tb]
\centering
\includegraphics[width=0.75\linewidth]{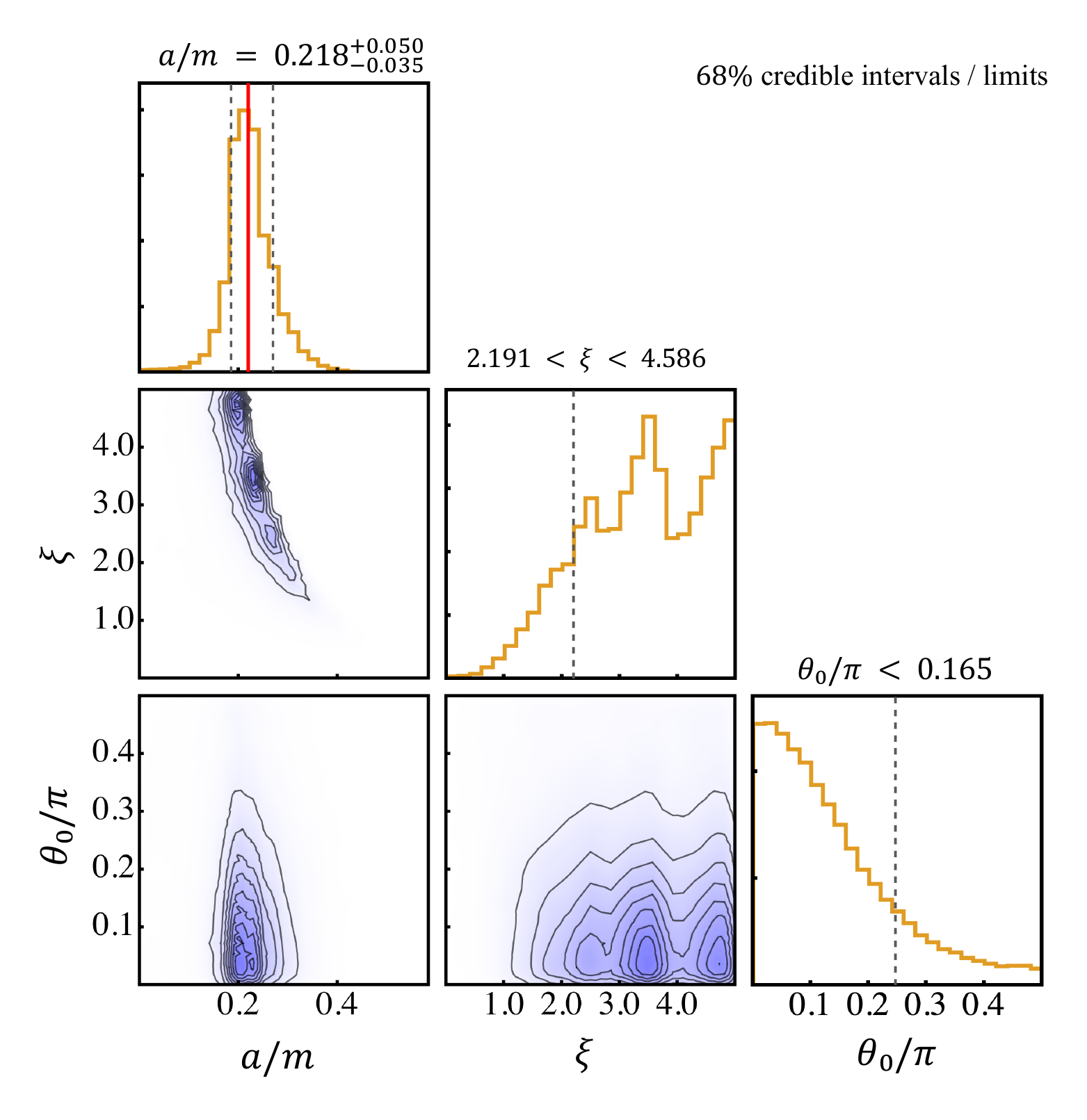}
\caption{Marginalized posterior distributions of $(a/m,\xi,\theta_0/\pi)$ for M87*.
Diagonal and lower-triangular panels show one- and two-dimensional posterior densities, respectively.
Contours indicate equal density; annotated intervals and upper limits correspond to 68\% credibility.}
\label{fig7}
\end{figure}

{
Figure \ref{fig7} presents the marginalized posterior distributions for M87*.
The median dimensionless spin and its 68\% equal-tailed credible interval are $a/m=0.218^{+0.050}_{-0.035}$, while the corresponding interval for the primary-hair parameter is $2.191<\xi<4.586$.
Both intervals are defined by the 16th and 84th percentiles of the respective marginal posterior distributions.
The inclination satisfies the 68\% one-sided upper credible limit $\theta_0/\pi<0.165$, corresponding to $\theta_0\lesssim29.7^\circ$.
Within the adopted inference framework, the posterior therefore favors relatively low spin and viewing directions closer to the rotation axis.
The two-dimensional posterior in the $(a/m,\xi)$ plane exhibits a pronounced anticorrelation, with larger hair parameters associated with smaller spins.
The elongated, curved distribution indicates a degeneracy between spin and primary hair.
This correlation reflects the combined effects of the observational likelihood and the horizon-restricted prior.
Although the hair posterior gives greater weight to larger values of $\xi$, it remains broad and retains appreciable density near the adopted upper boundary $\xi=5$.
The quoted hair interval should therefore be interpreted within this finite prior range, rather than as a bound independent of the adopted parameter domain.
\par}

\begin{figure}[tb]
\centering
\includegraphics[width=0.75\linewidth]{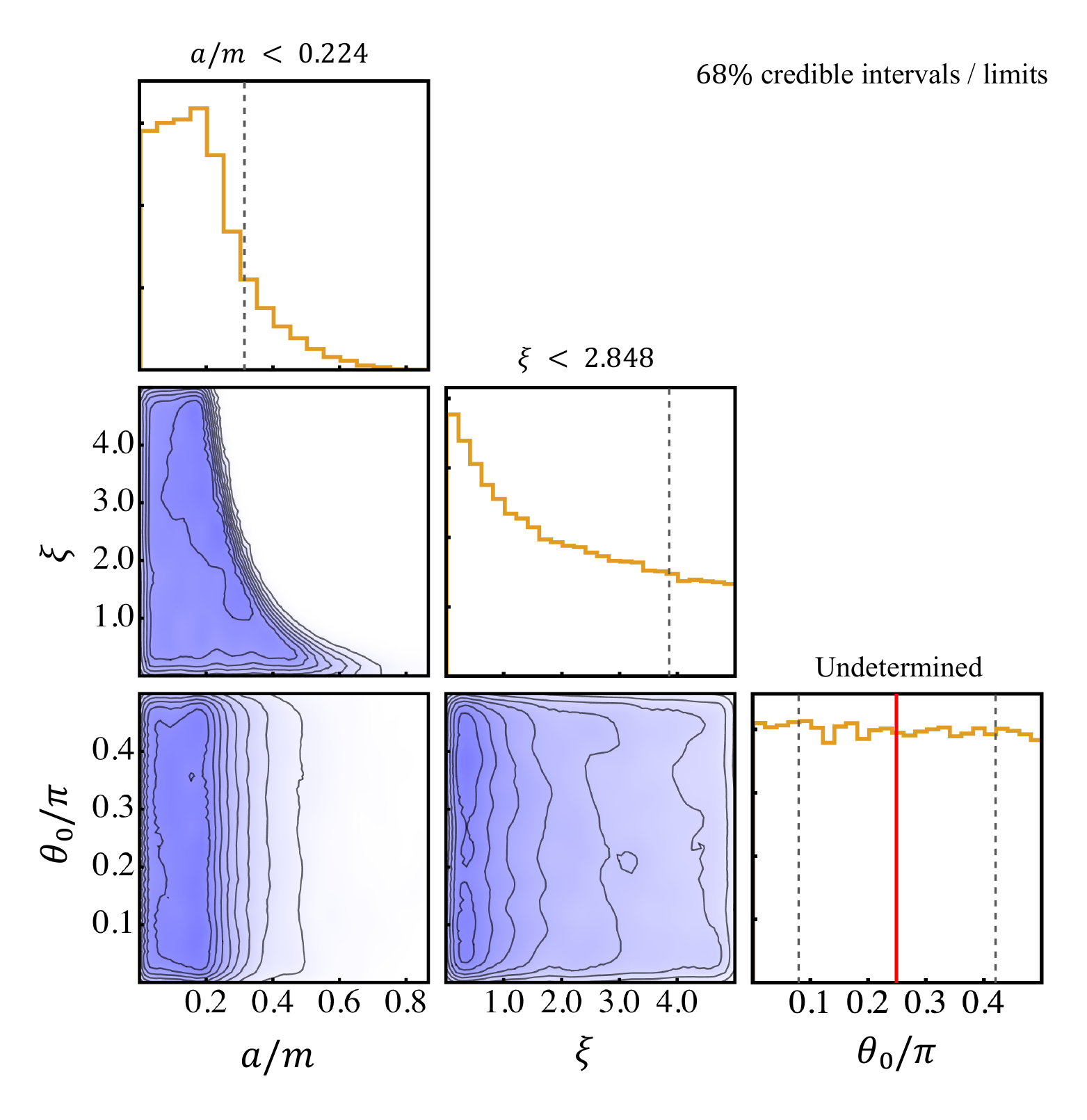}
\caption{Marginalized posterior distributions of $(a/m,\xi,\theta_0/\pi)$ for Sgr A*, with the same panel conventions as in Fig. \ref{fig7}.
The annotated upper limits correspond to 68\% credibility.}
\label{fig8}
\end{figure}

{
Figure \ref{fig8} shows the corresponding posterior distributions for Sgr A*.
The dimensionless spin and primary-hair parameter satisfy $a/m<0.224$ and $\xi<2.848$, respectively, at 68\% credibility, with both bounds defined as one-sided upper limits.
Unlike the relatively localized spin posterior for M87*, the Sgr A* distribution extends toward the nonrotating limit, while the hair posterior retains support near $\xi=0$.
The inclination posterior is approximately uniform over $0\leq\theta_0/\pi\leq0.5$, indicating that the adopted observables do not appreciably constrain the viewing direction under the chosen prior.
The joint distribution in the $(a/m,\xi)$ plane spans a broad region: posterior weight at larger $\xi$ is concentrated at smaller spins, whereas larger spins are associated with posterior weight at smaller $\xi$.
As for M87*, the horizon-restricted prior couples spin and primary hair, and the reported upper limits reflect its combined effect with the observational likelihood.
Taken together, Figs. \ref{fig7} and \ref{fig8} illustrate distinct posterior structures under the adopted observational inputs and prior.
M87* yields a more localized spin distribution and a preference for small inclination, while retaining a pronounced spin--hair degeneracy.
Sgr A* admits a broader family of configurations and leaves the inclination essentially unconstrained.
\par}

\par\vspace{0.8\baselineskip}
\section{Conclusion and discussion}\label{sec:conclusion}
{
We have investigated black hole shadows and gravitational lensing in an exact rotating solution with primary hair in Generalized Proca theory.
The hair modifies the geometry through a spin- and latitude-dependent correction that vanishes in the nonrotating limit.
Ray-traced images show a modest enlargement of the shadow for the parameter combinations considered, together with changes in the surrounding lensing pattern.
The area-equivalent radius and axis ratio characterize the shadow size and distortion, providing complementary information about primary hair, spin, and viewing inclination.
Applying these observables to M87* and Sgr A* yields distinct posterior distributions.
M87* favors relatively low spin and a viewing direction closer to the rotation axis, with a pronounced spin--hair anticorrelation.
The appreciable M87* posterior density near the adopted upper hair boundary also highlights the role of the prior range in its inferred hair interval.
Sgr A* yields upper limits on spin and hair while leaving the inclination essentially unconstrained.

A natural next step is to extend the present geometric analysis to emission from accretion flows \cite{Dhruv:2025grmhd}.
Studies of thin and thick disks in non-Kerr spacetimes have explored how spacetime parameters and accretion prescriptions shape the emitted radiation and observed images \cite{Mizuno:2018lxz,Chen:2011rx,Xu:2025iwg,Xu:2025jvk,Zeng:2021dlj,Zeng:2021mok,Gao:2023mjb,Li:2026anisotropicimages}.
Building on these approaches, the next step is to compute how the latitude-dependent deformation affects brightness asymmetries and higher-order lensed structures through radiative transfer in the primary-hair geometry \cite{Desire:2025multifrequency}.
Including polarized emission links these calculations to the magnetic structure and motion of the flow, following developments in photon-ring and accretion-flow polarimetry \cite{Himwich:2020msm,Wang:2025qpv,EventHorizonTelescope:2024rju,Hou:2025nearhorizon}.
The resulting maps can be used to predict interferometric signals, including photon-subring signatures on long baselines \cite{Johnson:2019ljv}.
These predictions would provide targets for the next-generation Event Horizon Telescope and the proposed Black Hole Explorer \cite{Johnson:2023ynn,Lupsasca:2024xhq}.
Combining the shadow scale, resolved brightness, and polarization in a joint Bayesian analysis \cite{Tiede:2026hibi} would extend existing image-based tests of black hole geometry \cite{EventHorizonTelescope:2022xqj,Ayzenberg:2023hfw,Chang:2025variability}, with plasma properties included in the inference to help distinguish primary hair, spin, and inclination.
This imaging program could be complemented by gravitational-wave tests of the underlying vector--tensor dynamics, including coupled gravitational--vector perturbations and mode-dependent spectral signatures \cite{Liu:2026cxs,Liu:2024Lv}, informed by studies of wave propagation, polarization, and cosmological perturbations \cite{Lai:2024einsteinvector,Dong:2025vectortensor,Lai:2026cosmovector}, as well as quasinormal modes and ringdown \cite{Liu:2026KRmodes,Liu:2023MOG}.
\par}

\acknowledgments The authors thank Dr. Wentao Liu for helpful discussions. J. Z. She acknowledges support from the Research Startup Fund of Tongling University (Grant No. 2025tlxyrc080), S. M. Wu acknowledges support from the National Natural Science Foundation of China (Grant Nos.12575056 and 12205133) and the LiaoNing Revitalization Talents Program (XLYC2503099).

\appendix

\section{Numerical construction of the deformed horizon}
\label{app:horizon_numerics}

\begin{figure*}[t]
\centering
\includegraphics[width=1\linewidth]{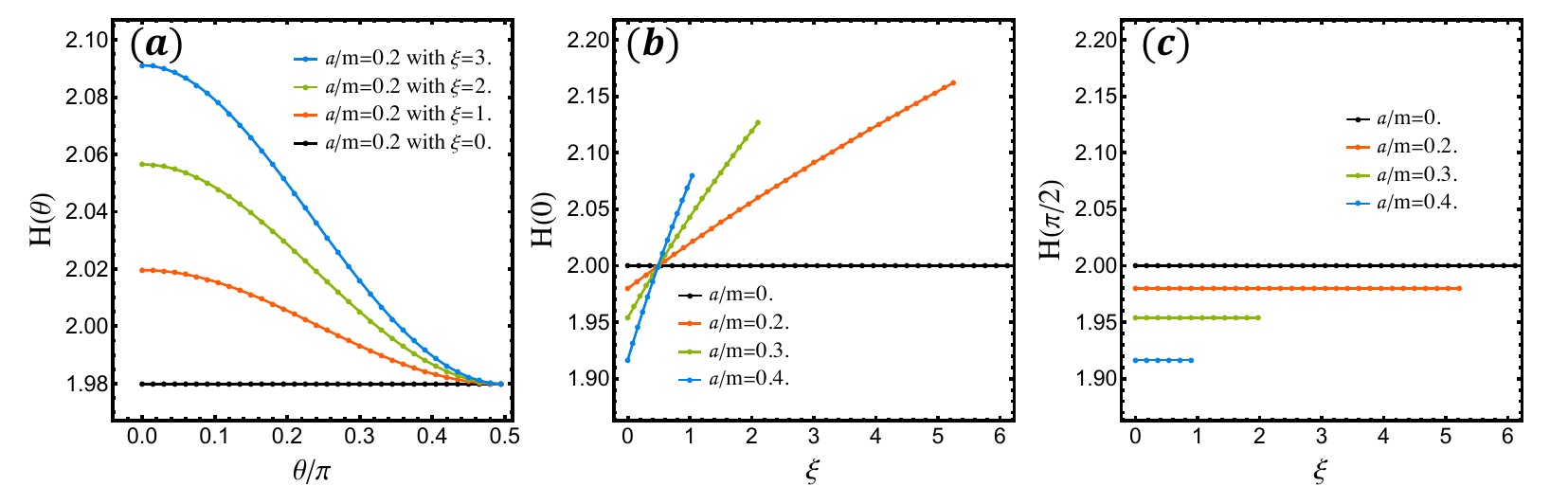}
\caption{Representative numerical outer-horizon profiles and their endpoint radii for $\xi\geq0$.
All radii are given in units of the ADM mass $m$.
(a) The dimensionless horizon profile $h(\theta)=H(\theta)/m$ at $a/m=0.2$ for several values of $\xi$.
(b) The polar radius $h(0)$ as a function of $\xi$ for several spins.
(c) The corresponding equatorial radius $h(\pi/2)$.}
\label{fig9}
\end{figure*}

We solve Eq. \eqref{deformedHorizon} on the reflection-symmetric interval $0\leq\theta\leq\pi/2$, restricting the numerical construction to $\xi\geq0$.
Since the horizon equation is invariant under reversal of the spin, it is convenient to work with the dimensionless quantities
\begin{equation}
\tilde{a}\equiv\frac{|a|}{m},
\qquad
h(\theta)\equiv\frac{H(\theta)}{m},
\qquad
\tilde\Delta(h,\theta)
\equiv\frac{\Delta(mh,\theta)}{m^2}.
\end{equation}
Equation \eqref{simpleMassFunction} then gives
\begin{equation}
\tilde\Delta(h,\theta)
=h^2+\tilde{a}^2
-2h\left\{
1+\xi\left[
\left(1+\frac{\tilde{a}^2\cos^2\theta}{h^2}\right)^2-1
\right]
\right\},
\end{equation}
so that the horizon equation takes the dimensionless form
\begin{equation}
\left(\frac{dh}{d\theta}\right)^2 +\tilde\Delta(h,\theta)=0.
\end{equation}
We begin at the equatorial plane, where the primary-hair correction vanishes.
The outer equatorial radius is therefore fixed by the Kerr root,
\begin{equation}
h_{\rm e}=1+\sqrt{1-\tilde{a}^2}.
\end{equation}
To build the reflection condition into the initial data, we introduce $u\equiv\pi/2-\theta$ and expand the outer branch as
\begin{equation}
h(u)=h_{\rm e}+c_{\rm e}u^2+\mathcal{O}(u^4).
\end{equation}
Substitution into the horizon equation yields
\begin{equation}
\begin{aligned}
c_{\rm e}^{\pm}
=\frac{-A_{\rm e} \pm \sqrt{A_{\rm e}^2 - 8B_{\rm e}}}{8},
\end{aligned}
\end{equation}
where $A_{\rm e}=2(h_{\rm e}\!-\!1) $ and $B_{\rm e}=-{8\xi\tilde{a}^2}/{h_{\rm e}}$.
For $\xi\geq0$, we select the larger root, which continuously approaches the constant Kerr solution $h=h_{\rm e}$ as $\xi\to0$.
Having fixed the regular outer branch in this way, we use its near-equatorial expansion to initialize the numerical integration.
Since the square-root form of the equation is degenerate at $u=0$, the integration cannot be started exactly at the equatorial plane; we therefore launch it at $u_0=0.01$.
The quadratic expansion supplies
\begin{equation}
h_0=h_{\rm e}+c_{\rm e}u_0^2,
\qquad
p_0=2c_{\rm e}u_0.
\end{equation}
We then apply three Newton corrections to $h_0$, keeping $p_0$ fixed, so that the original constraint is satisfied at the launch point,
\begin{equation}
p_0^2+ \tilde\Delta\left(h_0,\frac{\pi}{2}-u_0\right)=0.
\end{equation}

The outer branch is evolved across the interior interval according to
\begin{equation}
\frac{dh}{du}=F(u,h)
=\sigma\sqrt{-\tilde\Delta\left(h,\frac{\pi}{2}-u\right)},
\qquad
\sigma=\operatorname{sgn}(c_{\rm e}).
\end{equation}
The sign is held fixed while following this continuous branch.
A parameter point is discarded if the radicand becomes negative before the integration reaches the polar neighborhood.
We divide $u_0\leq u\leq\pi/2-u_0$ into $N=2000$ equal steps, with $\delta u=(\pi/2-2u_0)/N$, and use the fourth-order Runge-Kutta (RK4) method, whose update is
\begin{equation}
\begin{aligned}
k_1&=F(u_n,h_n),\\
k_2&=F\left(u_n+\frac{\delta u}{2},
h_n+\frac{\delta u}{2}k_1\right),\\
k_3&=F\left(u_n+\frac{\delta u}{2},
h_n+\frac{\delta u}{2}k_2\right),\\
k_4&=F\left(u_n+\delta u,
h_n+\delta u\,k_3\right),\\
h_{n+1}&=h_n+\frac{\delta u}{6}
\left(k_1+2k_2+2k_3+k_4\right).
\end{aligned}
\end{equation}

The RK4 evolution is continued until it reaches the neighborhood of the rotation axis, leaving a short endpoint interval that is not covered by the direct integration.
To complete the horizon profile across this interval, we construct a local expansion about the pole.
The polar radius $h_{\rm p}$ is determined as the outermost positive root of
\begin{equation}
h_{\rm p}^5-2h_{\rm p}^4+\tilde{a}^2h_{\rm p}^3
-4\xi\tilde{a}^2h_{\rm p}^2-2\xi\tilde{a}^4=0,
\end{equation}
which is equivalent to $\tilde\Delta(h_{\rm p},0)=0$.
A locally regular polar branch admits the even expansion
\begin{equation}
h(\theta)=h_{\rm p}+c_{\rm p}\theta^2
+\mathcal{O}(\theta^4),
\end{equation}
where the quadratic coefficient $c_{\rm p}$ is not fixed uniquely by local regularity.
Substituting this expansion into the horizon equation gives the two candidate values
\begin{equation}
c_{\rm p}^{\pm}
=\frac{-A_{\rm p}\pm\sqrt{A_{\rm p}^2-8B_{\rm p}}}{8},
\end{equation}
with
\begin{equation}
A_{\rm p}
=2h_{\rm p}-2
+\frac{4\xi\tilde{a}^2}{h_{\rm p}^2}
+\frac{6\xi\tilde{a}^4}{h_{\rm p}^4},\quad
B_{\rm p}
=\frac{8\xi\tilde{a}^2}{h_{\rm p}}
\left(1+\frac{\tilde{a}^2}{h_{\rm p}^2}\right).
\end{equation}
Thus $c_{\rm p}\in\{c_{\rm p}^{+},c_{\rm p}^{-}\}$.
Only real candidate values define locally regular polar branches.
At the last RK4 point, $u=\pi/2-u_0$ and hence $\theta=u_0$.
The value and the $u$-slope predicted by a polar branch are
\begin{equation}
h_{\rm p}^{\rm loc}=h_{\rm p}+c_{\rm p}u_0^2,
\qquad
p_{\rm p}^{\rm loc}
=\left.\frac{dh}{du}\right|_{\theta=u_0}
=-2c_{\rm p}u_0.
\end{equation}
The minus sign in the second relation follows from $du=-d\theta$.
We compare these quantities with the numerical endpoint value and slope and retain the real polar branch with the smallest combined mismatch.
This determines the polar endpoint branch used to complete the profile.

Finally, the RK4 data on the interior interval are represented by a cubic interpolating function.
The two short endpoint intervals omitted from the integration are filled with even-power local expansions through sixth order.
Their fourth- and sixth-order coefficients are fixed by matching the value and first derivative of the interior interpolant at $u=u_0$ and $u=\pi/2-u_0$.
This construction preserves
\begin{equation}
h'(0)=h'(\pi/2)=0
\end{equation}
and gives a continuously differentiable representation of $H(\theta)=m h(\theta)$ on the full interval.
The second hemisphere is then obtained by reflection.

The numerical results are illustrated in Fig. \ref{fig9}.
For $\xi=0$, the horizon reduces to the constant Kerr profile.
As $\xi$ increases, the primary hair produces an increasingly pronounced angular deformation: the polar radius increases, whereas all profiles retain the same equatorial endpoint.
This behavior is consistent with the fact that the primary-hair correction vanishes at $\theta=\pi/2$, so that $h(\pi/2)=h_{\rm e}$ is independent of $\xi$.
Panel (b) shows that, for nonzero spin, the polar radius grows with $\xi$.
The termination of each colored curve indicates the largest positive value of $\xi$ for which the numerical construction yields a regular outer-horizon profile at the corresponding spin.
In the nonrotating limit the solution remains $h(0)=2$.
The constant curves in panel (c), together with the common equatorial endpoints in panel (a), therefore provide a direct numerical check of the analytic equatorial boundary condition.

\end{document}